# In utero diffusion MRI: challenges, advances, and applications

Daan Christiaens PhD 1,2, Paddy J. Slator PhD 3, Lucilio Cordero-Grande PhD 1,2, Anthony N. Price PhD 1,2, Maria Deprez PhD 1,2, Daniel C. Alexander PhD 3, Mary Rutherford PhD 1,2, Joseph V. Hajnal PhD 1,2, Jana Hutter PhD 1,2

1. Centre for the Developing Brain, School of Biomedical Engineering & Imaging Sciences, King's College London, London, UK
2. Department of Biomedical Engineering, School of Biomedical Engineering & Imaging Sciences, King's College London, London, UK
3. Centre for Medical Image Computing and Department of Computer Science, University College London, London, UK

Correspondence to:

Jana Hutter, PhD

1st Floor South Wing, St. Thomas' Hospital

Westminster Bridge Road, SE17EH London

Tel: 0044-7510439022; FAX: 0044-2071889154

Jana.hutter@kcl.ac.uk

Funding Sources:

NIH Human Placenta Project, [1U01HD087202-01]

Wellcome Trust [201374/Z/16/Z]

Wellcome-EPSRC Center for Medical Engineering

EPSRC [N018702,M020533]

# In utero diffusion MRI: challenges, advances, and applications

Daan Christiaens[1,2], Paddy J. Slator[3], Lucilio Cordero-Grande[1,2], Anthony N. Price[1,2], Maria Deprez[1,2], Daniel C. Alexander[3], Mary Rutherford[1,2], Joseph V. Hajnal[1,2], Jana Hutter[1,2]

4. Centre for the Developing Brain, School of Biomedical Engineering & Imaging Sciences, King's College London, London, UK
5. Department of Biomedical Engineering, School of Biomedical Engineering & Imaging Sciences, King's College London, London, UK
6. Centre for Medical Image Computing and Department of Computer Science, University College London, London, UK

**Abstract**

In utero diffusion MRI provides unique opportunities to non-invasively study the microstructure of tissue during fetal development. A wide range of developmental processes, such as the growth of white matter tracts in the brain, the maturation of placental villous trees, or the fibres in the fetal heart remain to be studied and understood in detail. Advances in fetal interventions and surgery furthermore increase the need for ever more precise antenatal diagnosis from fetal MRI. However, the specific properties of the in utero environment, such as fetal and maternal motion, increased field-of-view, tissue interfaces and safety considerations, are significant challenges for most MRI techniques, and particularly for diffusion. Recent years have seen major improvements, driven by the development of bespoke techniques adapted to these specific challenges in both acquisition and processing. Fetal diffusion MRI, an emerging research tool, is now adding valuable novel information for both research and clinical questions. This paper will highlight specific challenges, outline strategies to target them, and discuss two main applications: fetal brain connectomics and placental maturation.

# 1. Introduction

The normal development of the fetus relies on numerous interrelated biological processes, transforming tissue within developing fetal organs in addition to the placenta across gestation  Deviations from these carefully synchronized developmental steps lead to consequences ranging in severity from developmental disorders to intra-uterine death[1]. Understanding and visualizing these changes increases knowledge about crucial biological events and can thus help support early and more accurate diagnoses, targeted medication or timed interventions, such as early delivery or intrauterine surgery.

Diffusion MRI (dMRI) has proved to be a powerful means to non-invasively explore and characterise tissue properties and has huge potential to contribute to these these intra-uterine insights, but comes with specific challenges. dMRI exploits the random motion of water, and specifically its hindrance by the surrounding cell membranes, to probe the tissue micro-architecture and thus to allow insights into tissue microstructure. Diffusion encoding gradients are employed, which can be varied in strength and timing to set sensitivity (b-value) and direction (b-vector). The b-value gives rise to different degrees of attenuation and thus informs on the restriction and size of structures. The b-vector can encode sensitivity to specific directions, hence allowing insights into tissue anisotropy.

Pulsed Gradient Spin-Echo (PGSE) Single-shot Echo-Planar-Imaging (ssEPI) is the most commonly employed technique. Each shot is composed of the diffusion preparation with a pair of monopolar gradients and the EPI readout train. Important parameters are echo time (TE) and repetition time (TR) — the latter defined as the time required to acquire all slices for one volume (with option to add a time pad if needed to allow extra magnetisation recovery or manage power demands on the scanner hardware from the use of strong gradients). Although dMRI is well established, notably for neuro-imaging applications[2] and oncology[3], its use for in utero applications is much less developed.

Ongoing improvements in techniques, in both acquisition and post-processing, are thus of crucial importance. In addition, depending on the studied in utero organ, the specific microstructural features of development call for a dedicated optimization of the diffusion encoding parameters (b-value/b-vector). dMRI data can also be used to quantify microcirculation of blood in the tissue (perfusion) using intra-voxel incoherent motion (IVIM) techniques[4]. IVIM MRI considers the signal to be a combination of two signal components: fast-attenuating and slow-attenuating. The fast-attenuating component is associated with blood perfusing in capillaries and is often termed "pseudo-diffusion". The slow-attenuating component is associated with water residing in tissue and is often termed "diffusion". Furthermore, interest is growing in combining relaxometry and dMRI both in modelling and in the acquisition. These efforts are typically aimed to disentangle the effects of both contrasts.

In the following, we will first discuss the specific challenges of in utero dMRI, secondly detail latest advances in strategies to deal with them, and lastly describe two large scale in utero dMRI studies in which our centre is involved in more detail: the developing Human Connectome Project (dHCP; www.developingconnectome.org) and the Human Placenta Imaging Project (PIP; www.placentaimagingproject.org).

## 2. Specific challenges of in utero dMRI

In utero diffusion-weighted imaging poses a unique and highly interdependent set of challenges:

Safety constraints: The in utero environment imposes enhanced safety constraints, in relation to specific absorption rate (SAR), peripheral nerve stimulation (PNS) and acoustic noise[5]. Low SAR is required to avoid tissue heating in the fetus, which has less capacity to lose heat due to the partially fluid filled intrauterine environment. In addition, the presence of the amniotic fluid itself causes increased radiofrequency magnetic field (B1) field variations leading potentially to regions of higher localised SAR within the fetus and maternal tissues[6]. The PNS level should be limited too, as increased maternal habitus increases likelihood that PNS may be experienced. Finally, limiting acoustic noise is required to protect the fetal hearing as additional hearing protection methods can not be applied in utero. Reduction of the noise to the fetus to <80dB(A) should be the goal. Therefore, taking the assumed sound attenuation of ~30dB(A)[7] from surrounding maternal tissues into account, the maximum permissible sound levels allowed should be reduced to <110dB(A) - which is significantly lower than what can be generated with dMRI and high gradient performance (>130 dB(A) on some 3T systems).

In practice, SAR is rarely the limiting factor for dMRI, due to the low density of RF pulses in conventional Spin-Echo (SE) sequences. However, both the low PNS and low acoustic noise requirements significantly constrain the acquisition, specifically the readout bandwidth. Both low PNS and low acoustic noise generally require a reduced slew rate, which increases the echo spacing and thus reduces the bandwidth of the EPI readout train. However, the longer readout time also renders the acquisition more susceptible to motion and geometric distortion at air-tissue interfaces.

In addition to critical safety aspects, maternal comfort is essential for a successful scan. Especially during the third trimester, women may find it uncomfortable to lay in the scanner for extended periods of time. Therefore, the acceptable scan time might be constrained, thereby limiting the number of diffusion encodings that can be obtained. Finally, maternal positioning is an area of important consideration. Both supine and left lateral tilt positions have been employed.

Region of interest: The location of the uterus inside the maternal abdomen contributes to another set of challenges: Independent of the region-of-interest (e.g. fetal brain or placenta), the maternal tissues determine the minimal achievable distance between coil and imaging region. Next, either a large Field-of-View or strategies for restricted FOV imaging are required. Both are suboptimal setups, resulting in lower SNR due to requirement of either a longer read-out train - thus increasing TE or required suppression strategies. This lower SNR in abdominal MRI is especially challenging for diffusion imaging, because dMRI in itself is also an inherently low-SNR modality as it deliberately attenuates the signal.

B0/B1 inhomogeneity and field strength: The presence of maternal tissues and amniotic fluid contributes to a challenging environment influencing both the B1 and static magnetic field (B0) inhomogeneity. Concretely for dMRI acquired with EPI techniques, which suffer in plane spatial distortion when there are variations in B0, susceptibility induced spatial B0 shifts that depend on the geometry of air-tissue interfaces, e.g., globally due to the mother's habitus and locally around bubbles in the maternal gut, require dedicated B0 shimming and post-processing strategies. B1 inhomogeneity, as mentioned above, is increased due to the presence of amniotic fluid around the fetus, leading to signal attenuation due to deviation from nominal flip angle. Strategies to avoid or correct these effects are paramount for accurate diffusion quantification. The degree of B0/B1 inhomogeneity is influenced furthermore by the employed field strength. While historically most fetal imaging was performed at 1.5T or even 0.5T, the SNR benefits of 3T are leading to an increasing number of sites employing 3T. All data shown in this review was acquired at 3T.

Motion: In utero dMRI is complicated by both maternal breathing and fetal motion, which may be abrupt, leading to inconsistent stacks of slices and requiring dedicated correction strategies. For dMRI, sudden motion during the diffusion weighting can lead to signal dropouts. Motion effects that are slow enough not to affect the acquisition of the slice, can lead to variable geometries. While the encoding is consistent in scanner space, the transformation to anatomical space results in encoding variation and thus variable diffusion encodings in every slice. Furthermore, motion can also lead to slice cross-talks artifacts (spin history effects) and therefore special emphasis needs to be placed on the slice acquisition ordering.

Finally, all the above mentioned challenges can influence and augment each other. For instance, subject motion leads to dynamically changing B0-induced distortions, and a reduced slew rate for noise protection increases distortions.

## 3. Recent advances addressing these challenges

### 3.1. Strategies to allow efficient and safe scanning

Conventional ssEPI sequences, successfully used for other diffusion acquisitions, have been employed for fetal diffusion, albeit with the restriction of low SAR, PNS and gradient performance. However, this approach often leads to decreased efficiency. Bespoke acquisition methods taking into account the specific challenges mentioned above, are required to improve efficiency and SNR under strict safety limits.

The efficiency of ssEPI depends on in-plane acceleration, the speed of k-space transversals (echo-spacing), the time for contrast weightings to evolve and the number of shots (equal to slices is single-band single-shot EPI is used) required to assemble each volume.

In PGSE sequences, the shortest echo time is achieved using unipolar gradients and then TE is typically constrained by the sum of the duration of the second diffusion gradient and the time from the start of the EPI readout to the center of k-space. Diffusion gradient duration can only be influenced by the gradient strength G for fixed maximal b-values. The time to the centre of k-space during readout can be reduced using parallel imaging, SENSE /Grappa and partial fourier methods. High partial Fourier factors increase sensitivity to eddy current artifacts (as the centre of k-space is closer to the diffusion gradients), higher SENSE factors decrease SNR, especially in combination with the suboptimal coil setup as discussed above. The EPI readout with its fast oscillating gradient is potent generator of acoustic noise. The echo-spacing (time between successive frequency encoded lines of k-space) defines the fundamental frequency of the EPI readout, which is the biggest determinant to acoustic noise. The relation between the EPI readout frequency and the acoustic noise generated is, however, non-linear and scanner-dependent. Therefore, system specific optimization is required.

Strategies taking these into account and optimizing the fundamental EPI read-out strategy can increase the efficiency as local minima at higher frequencies that nonetheless produce less noise can be exploited[8].

Multiband (MB) imaging (also known as simultaneous multi-slice (SMS)) is a technique that can be used to reduce volume TR and has seen rapid uptake in neuroimaging over the last few years. The total acquisition time is reduced by acquiring multiple slices simultaneously and using the parallel imaging techniques to unfold their profiles[9]. This has also been successfully applied in fetal ssEPI[10] but requires careful balancing between coil geometry and MB factor. The scope for acceleration using MB and in plane SENSE is more limited for in utero imaging than for many ex utero brain applications because the coils are more remote from the region of specific interest Further acceleration strategies are therefore of interest and include multiplexed (MP) EPI[11,12]. MP imaging reduces TR by acquiring, multiple slices by sequential RF pulses and then reading out consecutively within each k-space traversal. The time required for the increased encoding in the read direction and excitation pulses does, however, lead to a penalty in echo time (TE) and lengthens the overall EPI readout duration. Fetal imaging has restricted gradient switching due to acoustic noise and PNS limits, which increase echo spacing, thus may be a contender for multiplex applications[13].

### 3.2. Image denoising to overcome low SNR

The low SNR inherent to abdominal imaging makes advanced image denoising techniques of great interest for in utero dMRI. Recent developments in dMRI denoising exploit local redundancy across the dMRI series in patch-based singular value decomposition where a truncation (denoising) rank is determined based on the statistics of the noise spectrum[14]. In standard Gaussian noise, the spectrum follows a Marchenko-Pastur distribution. In low-SNR regimes, however, noise in magnitude images is Rician distributed and also spatially correlated due to SENSE and partial Fourier encoding. Although the obtained noise estimates and magnitude data can be modulated for Rician bias[14], access to the (demodulated) phase data opens new avenues for correcting noise in the complex domain, including accounting for the readout encoding[15].

### 3.3. Subject motion correction

Maternal respiration and fetal motion lead to inevitable motion artefacts in the imaging data. dMRI data sampled on regular parallel slices in scanner space, is sampled at scattered slice positions in subject space under the effects of motion. Correction strategies can either operate prospectively using navigator echoes that measure and adapt for subject motion during the scan, or retrospectively using image registration to estimate the unknown subject position at every time point during the scan.

Most conventional dMRI imaging uses 2D spin-echo EPI protocols in which motion can be assumed frozen during the readout of each slice or multiband slice-pack. Additional intra-slice motion will lead to signal dropouts that need to be rejected as outliers in correction strategies. As such, the slice or multiband excitation is the natural data unit in EPI, and the data should be seen not as a series of volumes, but as a collection of scattered slices. Motion correction then requires estimating the position of each slice with respect to the subject – or analogously the position of the moving subject at the readout time of each slice – using image registration, whilst simultaneously reconstructing the motion-corrected data on a regular grid in subject space. This process is known as slice-to-volume reconstruction (SVR) and is a common approach for anatomical imaging[16,17,18,19,20,21].

In addition, diffusion imaging has the special property that the sampled contrast is direction-dependent. Motion-induced subject rotation therefore also leads to scatter in the effective diffusion gradient encoding direction with respect to the subject. SVR techniques therefore need to be tailored for diffusion imaging to account for gradient reorientation at a slice level[22,23]. Doing so requires a means of predicting the dMRI contrast in unsampled diffusion encoding orientations, which can be done by representing the output image in a suitable q-space basis and/or by using appropriate priors[24,16]. Common q-space representations are the $2^{nd}$ order diffusion tensor[24,23] and the spherical harmonics (SH) basis for HARDI data[25], extended for multi-shell data[26]. All these techniques have in common that they treat motion correction as an inverse problem, where an uncorrupted data representation and slice-level motion traces are jointly estimated from the acquired scattered slice data.

For fetal brain imaging, motion can be assumed rigid and the required image registration can be constrained as such. Placenta imaging will generally be subject to non-rigid motion due to maternal respiration and uterine contractions, which may require non-linear registration methods to be incorporated into SVR techniques.

## 3.4. Dynamic adapted distortion correction

In addition to subject motion, EPI is also subject to severe image distortion due to B0 susceptibility arising largely from presence of air-tissue interfaces. The local in utero environment is composed of tissues of similar susceptibilities, however gas in maternal gut induces variations of the B0 field that is smooth inside the uterus and obeys a Laplacian constraint[27]. When imaging at 1.5T, the distortion field is generally treated as independent of maternal and/or fetal motion, so correction strategies have relied on using pre-acquired field maps or image driven registration that assumes a static B0 field[24,27]. However, susceptibility induced distortion increases linearly with field strength, and at 3T variability of B0 field with maternal breathing becomes a problem. Moreover, HARDI diffusion scans are lengthy, resulting in change of the B0 field over time due to displacing air-tissue interfaces, particularly in the presence of any gas bubbles in the maternal gut. Correcting EPI distortion in in utero imaging therefore requires a dynamic approach in which the B0 field map is estimated at every time point throughout the acquisition. In addition, incomplete saturation of the maternal fat can lead to unwanted fold-over artifacts if not considered as part of the initial B0 shimming procedure[28].

Dynamic field mapping has been achieved with double spin echo sequences where the second readout is traversed with reversed phase encoding. Assuming no subject motion between both echoes, each will have identical but opposite distortion. Symmetric nonlinear image registration can then be used to retrieve the field in each slice[29]. Nevertheless, the low SNR of long-TE dMRI in the 2nd echo can make such approach challenging at higher b-values, limiting the precision of dynamic field mapping. Therefore, we have proposed to interleave high-b and low-b sensitized slices within every volume in order to obtain the same data with temporally consistent sampling of the low-b slices[30]. These low-b slices provide sufficient SNR to estimate the local field map, which is subsequently interpolated in time to predict the dynamic distortion throughout the duration of the acquisition protocol.

An alternative acquisition strategy to deal with the aforementioned dynamic distortion effects is to estimate the distortion based on the phase evolution between consecutive spin and gradient echoes. Either due to relatively long T1 in the fetus or SAR constraints, some dead time may be available after finalizing the spin echo readout and applying the next excitation. This time can be used to acquire a field echo with same inter-echo spacing than the spin echo so that both reconstructed echoes present the same level of distortion. Then, by complex subtraction of the spin from the field echo, the components due to both bulk and physiological motion during the application of the diffusion gradients can be removed from the field echo phase. The resulting accumulated dephasing linearly depends on the B0 with a slope given by the difference between the field and spin echo times. After appropriate phase unwrapping, distortions can be corrected in a similar manner to phase-based corrections for fMRI[31]. However, whereas field maps can be directly inferred from the gradient echo images acquired for fMRI, the use of spin echoes for dMRI removes the field dependent phase in the primary images, but adds a motion dependent phase. Use of a combined Spin And Field Echo (SAFE) technique[32], which allows B0 field estimation from the phase difference between consecutive spin echo and gradient echo readouts (Fig.2d) after correcting for motion induced phases, has shown potential in enabling corrections at lower contrast to noise ratios than magnitude-based methods. However, these corrections remain limited by dropouts in the gradient recalled echo, which is collected second, so has low SNR and introduces a long dephasing time.

## 4. Examples

### 4.1. Fetal brain imaging

4.1.1. Background

Advanced insight into brain development in normal and at risk fetuses can greatly impact our understanding of neurodevelopmental disorders. The fetal brain undergoes a series of formative growth processes during the third trimester of pregnancy, which are vital for normal development. Any alterations or delays in these processes is likely to impact the subsequent neurodevelopment and quality of life of the child. MRI can provide unique, non-invasive insight into brain development during this critical phase, allowing both genetic and environmental influences to be assessed. In utero MRI can therefore offer earlier and more accurate diagnosis of neurodevelopmental complications in the fetal brain and may be used to predict subsequent outcomes for the child.

The developing Human Connectome Project (dHCP, www.developingconnectome.org) is an ERC-funded effort to obtain comprehensive imaging and collateral data in over 1000 neonates and fetuses, enabling the detailed study of brain development during this critical perinatal period. The imaging data includes diffusion MRI, functional MRI and anatomical MRI, all acquired at 3T. The overarching aim of the project is to create a comprehensive map of structural and functional brain connectivity (the human connectome) in early development. To this end, the project is also developing pipelines for reconstruction and motion correction, cortical parcellation, longitudinal analysis and brain atlas construction. All data, including outputs of standard processing pipelines, will be made available and hence offer a valuable, open resource for the neuroscience community.

Diffusion MRI plays a unique role in its capacity to probe white matter microstructure and fibre orientation. When integrated across the brain, the local axon orientation provides vital information to track white matter pathways and map the structural connectome. Fetal and neonatal dMRI thus provide valuable information on early white matter development[33] and are instrumental to the aims of the dHCP.

4.1.2. Imaging and processing pipeline

The low SNR inherent to abdominal diffusion imaging imposes an upper limit on the achievable b-value. The diffusion encoding in the fetal dHCP protocol[34] was therefore designed to consist of b=0s/mm²,b=400s/mm² and b=1000s/mm² shells, optimized for maximal contrast-to-noise . All data is acquired on a 3T Philips Achieva scanner using a 32-channel cardiac coil. The dMRI protocol uses the SAFE sequence[32] that combines spin echo and field echo readouts to facilitate dynamic distortion correction as described above. The acquisition is set at 2mm isotropic resolution with MB factor 2 and SENSE factor 2, planned in pure transverse plane centered on the fetal brain but irrespective of its pose. This ensures that acceleration is always supported by the combination of coil properties and image geometry. Given the available acceleration this enables acquiring 141 unique diffusion encoding gradients in 13:45min.

The processing pipeline integrates image denoising based on random matrix theory as described in Section 3.2[15], dynamic distortion correction based on the phase evolution between spin and field echoes as described in Section 3.4, and retrospective slice-level motion correction for multi-shell data[18]. The motion correction leverages a data-driven q-space representation in spherical harmonics and a radial decomposition (SHARD) for multi-shell data[26], and also incorporates slice outlier rejection. Following motion and distortion correction, the data is aligned in a self-consistent reference frame, as illustrated in Fig. 3 by the improved image alignment both within and across shells., In this self-consistent reference, the local signal can then be fitted with advanced signal and tissue models. Figure 4 shows an example fetal dMRI dataset after processing and its local tissue orientation distribution function (ODF), estimated in the output using constrained spherical deconvolution (CSD) in the outer shell[35]. The image shows how developing white matter tracts such as the corpus callosum and corticospinal tract are successfully reconstructed. Additionally, the figure reveals clear radial tissue organisation in cortical gray matter typical in the early developing brain.

## 4.2. Placental imaging

4.2.1. Background

The placenta is the crucial link between mother and fetus during pregnancy, providing oxygen and nutrients to the fetus and removing waste products. It is an organ with unique characteristics; having both fetal and maternal blood circulations which are separate and non-mixing. The placenta is implicated in many major pregnancy complications, such as pre-eclampsia (PE)[36] fetal growth restriction (FGR)[37] over invasive placentation. In these complications placental dysfunction may occur weeks before symptoms present in the fetus or mother, providing a potential early warning sign of disease. New MRI techniques for placental imaging hence have the potential for prediction or earlier diagnosis of pregnancy complications, in addition to assessing the degree of invasion in cases of invasive placentation.

Diffusion MRI techniques provide numerous benefits over ultrasound, such as a larger field of view enabling whole-organ assessment, higher contrast and the ability to provide functional information. Figure 5 shows examples parameter maps from a typical placenta dMRI scan.

Placental dMRI can differentiate placentas in pregnancies complicated by FGR and PE from healthy controls by estimating microstructure and microcirculatory properties. A majority of placental dMRI studies in the literature have utilised intravoxel incoherent motion (IVIM) MRI[38,39,40,41,42,43,44,45] which yields information on perfusion and diffusion regimes separately.

However, despite the promise shown, dMRI is not yet part of clinical practice. The development of novel dMRI scanning techniques specific to the unique structural and functional characteristics of the placenta may be necessary to overcome this in addition to user friendly clinically relevant post acquisition analysis.

4.2.2. Combination of dMRI with relaxometry in the placenta

Another MRI technique which shows much promise in the placenta is T2* relaxometry. This technique utilises the sensitivity of the transverse relaxation time to tissue structure and function. The T2* property most relevant for placental scanning is its sensitivity to oxygen. Multiple studies have shown that placenta T2* decreases in FGR[45,44], likely reflecting decreased oxygenation levels.

Although dMRI and relaxometry both show sensitivity to placental dysfunction, the inherent correlations between these techniques imply that measuring them separately is not ideal. An alternative approach is to measure both at the same time, aiming to disentangle the effects of each contrast. One such study combines T2 relaxometry with dMRI in the placenta, allowing separation of maternal and fetal blood circulations[46]. The acquired data consisted of multiple diffusion-prepared spin echos with varying b-value and echo time on six in-vivo human placentas. The scans were analysed with the DECIDE model[46], which accounts for three separate tissue compartments: fetal blood (characterised by high diffusivity and long T2), maternal blood (low diffusivity, long T2), and tissue (low diffusivity, short T2). DECIDE fitting allows estimation of multiple parameter maps: T2 of fetal and maternal blood, maternal to fetal blood volume ratio, and — when combined with fliterature observations — the fetal blood oxygen saturation.

Although dMRI methods such as intravoxel incoherent motion (IVIM) MRI are sensitive to blood flow, this only applies to incoherent (i.e. randomly orientated) flow. A recent study sought to address this by combining dMRI with measurements of coherent blood flow within the placenta. Net blood flow velocities were estimated in three orthogonal directions and mapped throughout the placenta. By combining these with IVIM parameter maps the authors measured the relative contributions of coherent and incoherent flow throughout the placenta.

4.2.3. Anisotropic IVIM

An inherent property of the IVIM model is that it treats perfusion (and diffusion) as isotropic. This assumption may not hold in the placenta, as the vascular tissue types have a coherent architecture. For this reason models which account for anisotropy in perfusion and diffusion regimes — termed anisotropic IVIM — have been explored for the placenta. A rich dMRI protocol was deployed to assess the types of model that the placental dMRI signal potentially supports. Our conclusion is that anisotropic IVIM models describe the placental dMRI signal better than standard models such as IVIM, ADC and DTI[47]. This observation motivated the development of a bespoke dMRI sequence for placental scanning.

4.2.4. Combined diffusion-relaxometry

As mentioned earlier, T2* relaxometry and dMRI have both shown promise for assessing pregnancy complications, but inherent correlations between these contrasts mean that measuring them separately is not optimal. We therefore undertook combined diffusion-relaxometry scanning in the placenta. This utilised a novel MRI acquisition strategy, termed ZEBRA[48], which allows sampling of multiple echo times and diffusion encodings within a single repetition time (TR). By appending multiple gradient echoes after a diffusion-prepared spin echo sequence, T2* and diffusion contrast is acquired. The data, including both healthy controls and various pregnancy complications, was analysed by fitting a continuum model, which assumes a distribution of diffusivity and T2* values[49]. This yielded estimates of the joint 2D ADC-T2* spectra, i.e., an estimate of the 2D distribution of ADC and T2* values in the placenta. We observed multiple separated peaks in these spectra, potentially reflecting multiple tissue microenvironments characterised by their T2*, ADC values (Figure 6). These ADC-T2* spectrum also show clear qualitative separation between control participants and those with pregnancy complications. Although we emphasise the small sample number, this highlights a potential advantage of combining dMRI with other contrasts: it provides a higher dimensional space for separating control and disease cases. Further work will focus on acquiring these richer types of dMRI data on cohorts of specific pregnancy complications.

## 5. Discussion and Conclusion

Diffusion MRI provides a unique window of observation into in utero tissue microstructure and development, with wide-ranging possibilities to target the contrast specifically to the research question and/or clinical application of interest. However, technical challenges related to subject motion, low SNR and MRI safety limit the success of conventional dMRI techniques in fetal imaging. Novel developments in both acquisition and post-processing have therefore been instrumental to overcoming the challenges of in utero dMRI. The recent advances in efficient scanning, motion and distortion correction, and noise suppression discussed in this review illustrate how substantial improvements can be achieved with innovations in all parts of the pipeline from dMRI acquisition to artifact correction and modelling. These innovations will facilitate more advanced analyses of in utero tissue microstructure that can ultimately lead to essential new insights in fetal development.

This paper focused on the two main applications of in utero diffusion MRI: fetal brain and placenta imaging. Fetal brain dMRI offers promising potential for noninvasive diagnosis of neurodevelopmental disorders[50,51]. Placenta dMRI can provide vital insight into oxygenation processes in the crucial link between mother and fetus. Nevertheless, other applications are emerging in research settings and offer interesting potential for more elaborate clinical studies. Fetal heart dMRI was successfully demonstrated in the sheep model[52] and ex vivo[53]. Other studies have explored antenatal kidney assessment with dMRI[54] and fetal lung maturation over gestation[55,56].

Owing to their extensive technical challenges, most clinical applications of in utero dMRI are still in their infancy. However, broader availability of novel acquisition and postprocessing techniques can accelerate clinical translation. dMRI is increasingly part of clinical MRI exams for assessing fetal brain development and injury[57,58,59]. Diffusion imaging has also been included in the assessment of invasive placentation in clinical settings[60,61].

Ongoing and future developments in MRI acquisition and postprocessing will continue to advance in utero dMRI translation into clinical research and practice. Acquisition developments include bespoke fetal coils, and closely related further improvement of acceleration techniques. Quiet scanning techniques designed to reduce acoustic stimulation can also improve patient comfort and scanning efficiency in the future. Regarding post-processing, a broad range of techniques demonstrated either in neonatal diffusion imaging or in fetal anatomical imaging can be ported and tailored to the specific challenges in fetal dMRI. For example, we anticipate recent developments toward machine learning techniques for automatic localization and pose correction of the fetal brain[62] to be adapted to fetal dMRI. Similarly, longitudinal atlases of normal tissue maturation of the neonatal brain over gestation[63,64,65] are being extended to fetal data, in line with similar developments in anatomical MRI[66]. These will ultimately allow assessing developmental stages at the group and subject level. Finally, integration with fetal cortical parcellation atlases can enable deriving network matrices of structural connectivity in the fetal brain.

Figures:

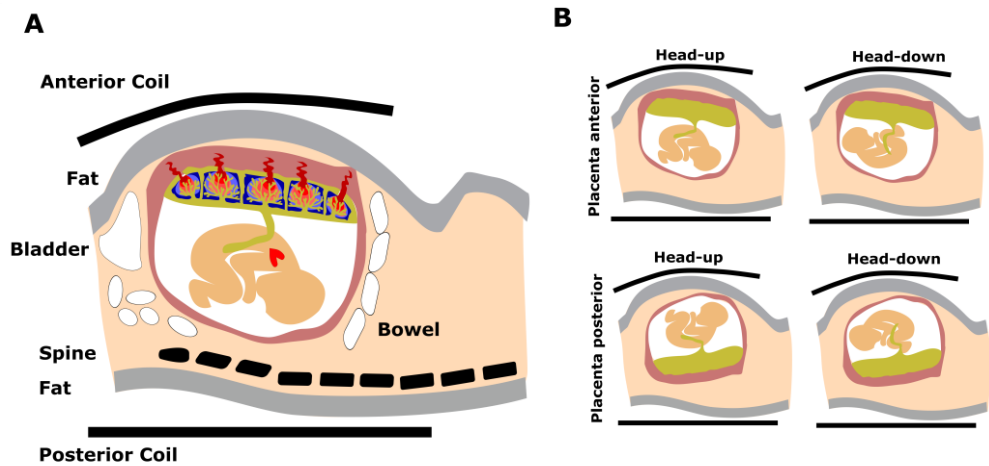

Figure 1: Schematic representation of the scan setup for an in utero dMRI scan. A. Overview and B. Variations in placental and head locations.

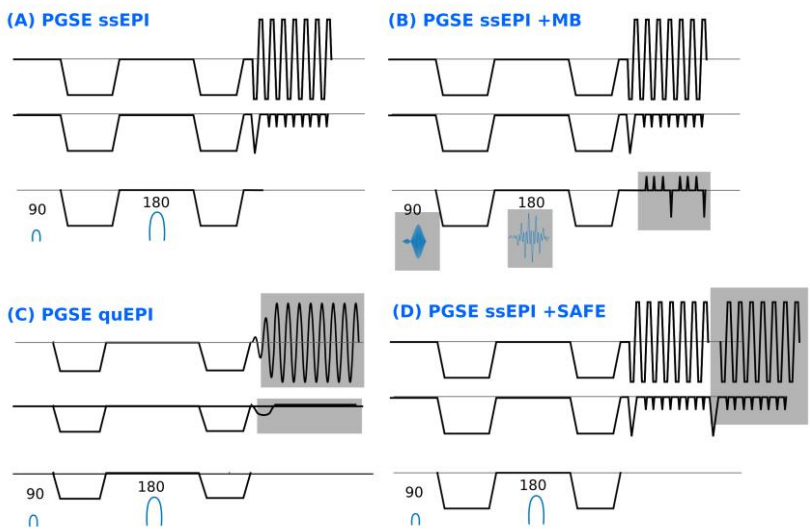

Figure 2: PGSE ssEPI diffusion-weighted MRI sequence with discussed improvements. A basic sequence, B sequence with MB acceleration, C sequence with quiet EPI read-out and D sequence with SAFE.

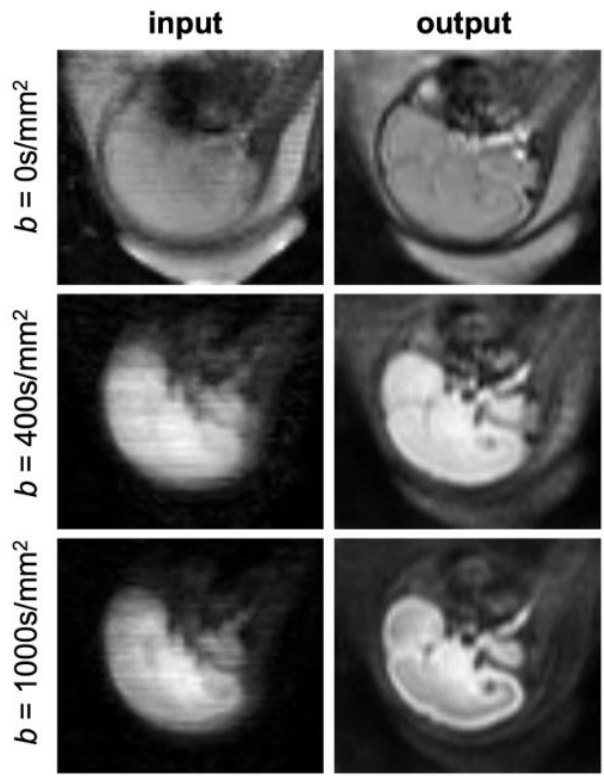

Figure 3: Motion correction in the fetal dMRI data of the developing Human Connectome Project. The left column shows the mean dMRI contrast in each shell before motion correction. The right column shows improved alignment after motion correction.

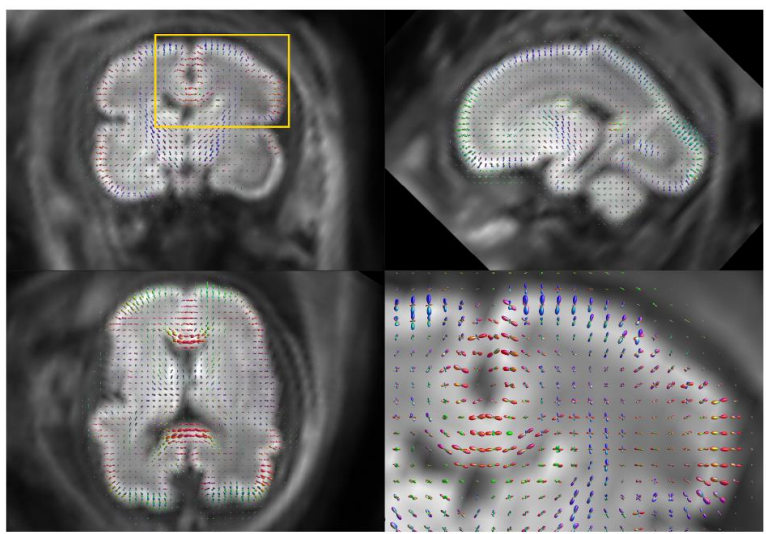

Figure 4: Example fetal dMRI dataset (32 weeks gestational age) of the developing Human Connectome Project, after processing for motion and distortion correction. The image shows the local fibre orientation distribution in each voxel, estimated using constrained spherical deconvolution. The background image is the mean b=1000s/mm² image after motion correction.

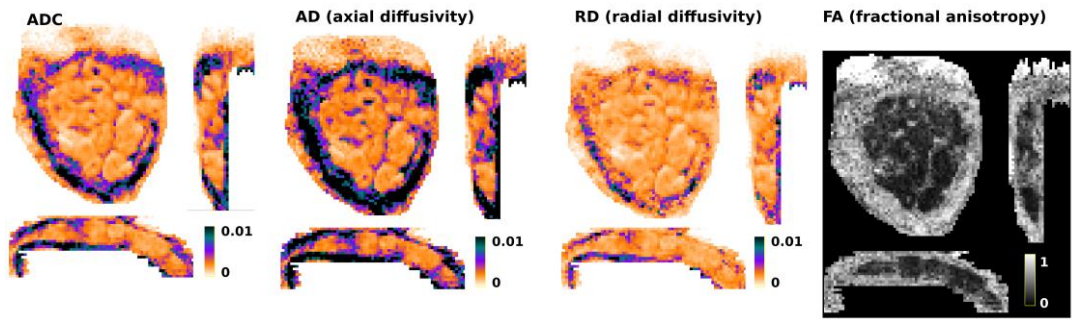

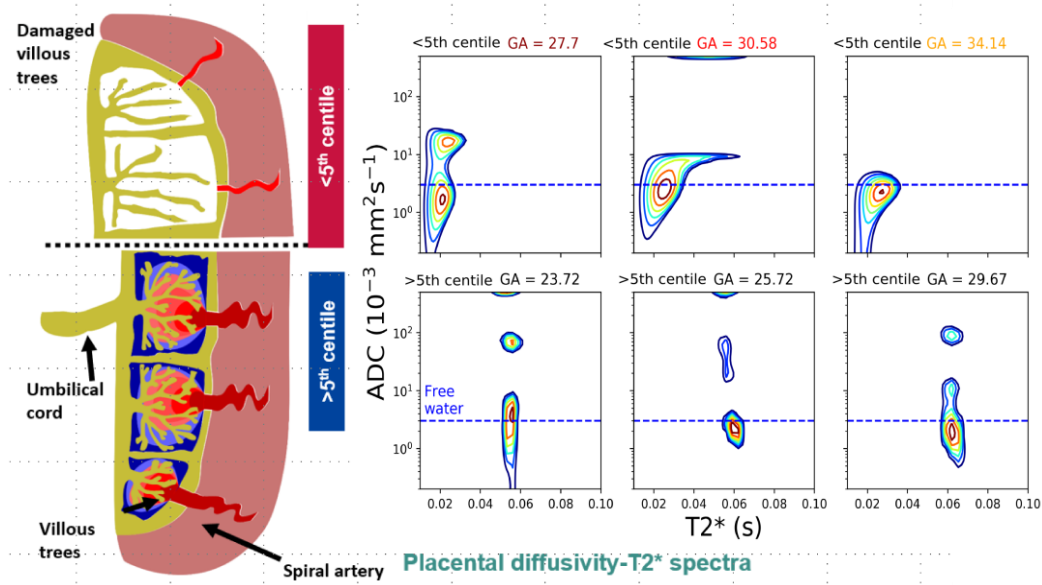

Figure 5: Example placental dMRI dataset (30 weeks gestational age) of the Placental Imaging Project. The image shows the Apparent Diffusion Coefficient (ADC) map, the Axial diffusivity, the radial diffusivity and the fractional anisotropy map in coronal, sagittal and transverse orientation.

Figure 6: Example placental T2*-dMRI results from the Placental Imaging Project. (Left) a schemata of the important microstructural elements in the placenta both in complicated pregnancies (top) and healthy pregnancies (bottom). (right) The image shows the results of three placentas of fetuses with eventual birthweight below the 5th centile and three with weight above the 5th centile together with the underlying biological hypothesis which might explain the differences.